\documentclass[12pt]{article}
\usepackage{graphics}
\usepackage{url}
\usepackage{color}
\usepackage{cite}

\newcommand{\gsim}{\lower.7ex\hbox{$\;\stackrel{\textstyle>}{\sim}\;$}}
\newcommand{\lsim}{\lower.7ex\hbox{$\;\stackrel{\textstyle<}{\sim}\;$}}

\newcommand{\beqn}{\begin{eqnarray}}
\newcommand{\eeqn}{\end{eqnarray}}
\newcommand{\tch}{ $t\to c H$}
\newcommand{\eebs}{$e^+e^-\to b \bar{s}$}
\newcommand{\eebsh}{$e^+e^-\to b \bar{s} H$}
\newcommand{\ppbs}{$pp\to b \bar{s}$}
\newcommand{\Ovdftwo}{${\cal O}(v^2/f^2)$}

\begin{document}
\title{Flavor changing top quark decay and bottom-strange production
in the littlest Higgs model with T-parity }
 \vspace{3mm}
\author{
Zhou Ya-Jin$^{a}$\footnote{zhouyj@sdu.edu.cn}, Hou Hong-Sheng$^b$\footnote{hshou@hznu.edu.cn}, Sun Hao$^c$\footnote{haosun@mail.ustc.edu.cn}\\
{\small $^a$ School of Physics, Shandong University, Jinan Shandong 250100, P.R. China} \\
{\small $^b$ Department of Physics, Hangzhou Normal University, Hangzhou Zhejiang 310036, P.R.China} \\
{\small $^c$ School of Physics and Technology, University of Jinan,
Jinan Shandong 250022, P.R.China} }

\date{}
\maketitle \vskip 12mm
\begin{abstract}
Flavor changing effects on the processes \tch, \eebs, \eebsh~ and
\ppbs~ in the LHT model are investigated in this paper. We calculate
the one-loop level contributions from the T-parity odd mirror quarks
and gauge bosons. The results show that the top quark rare decay
\tch~ in the LHT model can be significantly enhanced relative to
that in the SM. The $b\bar{s}$ production at linear colliders in the
LHT model can enhance the SM cross section a lot and reach 0.1 fb in
some parameter space allowed in the experiment. But the heavy gauge
boson and mirror fermion loops have small contribution to the
processes \ppbs~ and \eebsh. So the LHT effect on \eebs~ might be
detected at future linear colliders, while it's too small to be seen
for the \eebsh~ and \ppbs~ processes at future linear colliders and
LHC.
\end{abstract}


\vfill \eject \baselineskip=0.32in
\makeatletter      
\@addtoreset{equation}{section}
\makeatother       
\section{Introduction}
\par
The Flavor Changing Neutral Currents (FCNC) couplings play an
important role in searching for new physics beyond the Standard
Model (SM) for the following reasons: they are forbidden at tree
level and suppressed by the GIM mechanism\cite{Glashow:1970gm} at
one loop level in the SM, on the other hand, many new particles
appear in the loops in new physics models beyond SM, which may
enhance the flavor changing transitions\cite{BSM.FC}. Furthermore
searching for FCNC is one of the main goals of the high energy
colliders.

\par
The Little Higgs mechanism\cite{LH-origin,LH-review} offers a
solution to the hierarchy problem without fine tuning. The most
compact implementation of the Little Higgs mechanism is known as the
Littlest Higgs model (LH)\cite{N0206021,H0301040}. But the original
Little Higgs models suffer strong constraints from electroweak
precision data\cite{constraints}. To solve this problem, a ${\cal
Z}_2$ discrete symmetry named ``T-Parity" is introduced in
Ref.\cite{LHT, I0409025}, by which dangerous diagrams with the tree
level exchange of heavy neutral gauge bosons are forbidden. The
Littlest Higgs model with T-parity (LHT) requires the introduction
of ``mirror fermions" for each SM fermion doublet. The mirror
fermions are T-odd and can obtain large masses. These mirror
fermions and the heavy gauge bosons appear in loop diagrams can
induce new contributions to quark FCNC processes.

\par

FCNC effects are usually studied via decay mode, such as rare $B, K,
D$ meson decay, top quark rare decay, Higgs and Z decay, etc.
Besides via decay modes, the flavor changing vertices
$bsV$($V=\gamma,Z$) can also be investigated via bottom-strange
associated production. The \eebs ~process in the SM have been
studied in \cite{Huang:1999yu}, and the results show that the cross
section is larger than that of the $t\bar{c}$ production. We checked
these two processes in the LHT model, and found that this character
is kept. We will study the flavor changing effects via $b\bar{s}$
and $b\bar{s}H$ productions at linear colliders in the LHT model.
Since the phase space of $b\bar{s}$ production at the large hadron
collider (LHC) is very large, we will investigate this process as
well and looking forward to get large contribution. On the other
side, the top quark plays a special role in FCNC phenomenology due
to its heaviness. The decay mode \tch ~is forbidden at tree level,
and its decay width is about $10^{-14}$ at one loop level in the
SM\cite{Mele:1998ag}. In the MSSM the width could be enhanced to
$10^{-5} \sim 10^{-4}$\cite{tch.mssm}. In Ref.\cite{Tabbakh:2005kf}
the \tch ~process is studied in the LH model, and the authors found
that the branch ratio is at most of the order $10^{-12}$. With more
mirror particles in the loops and less constraints on the
parameters, \tch ~process in the LHT model would have much larger
decay width than in the LH model.

\par
In this paper we investigate the flavor changing effects on the
following processes: \tch, \eebs, \eebsh ~and \ppbs. Since the final
states $b\bar{s}$ and $\bar{b}s$ are undistinguishable in the
experiment, we need to calculate the two individual processes
$b\bar{s}(H)$ and $\bar{b}s(H)$ and sum their cross sections up. We
found that the two individual processes have the same cross
sections, so we only give the results for $b\bar{s}(H)$ final state,
the total cross sections can be obtained by doubling the
$b\bar{s}(H)$ ones. The paper is organized as follows: In Sec.II we
have a brief review of the LHT model. In Sec.III we give the
analytical and numerical calculations. Finally a short summary is
given.

\section{Brief review of the model}
\par
Here we briefly review the structure of the LHT model relevant to
our analysis, and the detailed description can be found in the
literature\cite{Hubisz:2004ft, Hubisz:2005bd}.

\subsection{Gauge and Higgs sectors}
\par
Gauge and Higgs sectors of the littlest Higgs model are described as a nonlinear
$\sigma$ model with the spontaneous global symmetry breaking from SU(5) to SO(5)
with scalar fields. From the SU(5)/SO(5) breaking, there arise 14 Goldstone bosons
which are described by the ``pion" matrix $\Pi$, given explicitly by
\beqn
\label{Pi}
\addtolength{\arraycolsep}{3pt}
\renewcommand{\arraystretch}{1.3}
\Pi=\left(
\begin{array}{ccccc}
-\frac{\omega^0}{2}-\frac{\eta}{\sqrt{20}} &
-\frac{\omega^+}{\sqrt{2}} &
  -i\frac{\pi^+}{\sqrt{2}} & -i\phi^{++} & -i\frac{\phi^+}{\sqrt{2}}\\
-\frac{\omega^-}{\sqrt{2}} &
\frac{\omega^0}{2}-\frac{\eta}{\sqrt{20}} &
\frac{v+h+i\pi^0}{2} & -i\frac{\phi^+}{\sqrt{2}} & \frac{-i\phi^0+\phi^P}{\sqrt{2}}\\
i\frac{\pi^-}{\sqrt{2}} & \frac{v+h-i\pi^0}{2} &\sqrt{4/5}\eta &
-i\frac{\pi^+}{\sqrt{2}} & \frac{v+h+i\pi^0}{2}\\
i\phi^{--} & i\frac{\phi^-}{\sqrt{2}} & i\frac{\pi^-}{\sqrt{2}} &
-\frac{\omega^0}{2}-\frac{\eta}{\sqrt{20}} & -\frac{\omega^-}{\sqrt{2}}\\
i\frac{\phi^-}{\sqrt{2}} &  \frac{i\phi^0+\phi^P}{\sqrt{2}} &
\frac{v+h-i\pi^0}{2} & -\frac{\omega^+}{\sqrt{2}} &
\frac{\omega^0}{2}-\frac{\eta}{\sqrt{20}}
\end{array}
\right).
\eeqn
where it consists of a doublet $H$ and a triplet $\Phi$ under
the unbroken $SU(2)_L \times U(1)_Y$ group which are
given by
\beqn
 H= \left(
\begin{array}{c}
-i\frac{\pi^+}{\sqrt{2}} \\
\frac{v+h+i\pi^0}{2} \end{array}\right),~~~~~~
\Phi=
\left(\begin{array}{cc} -i\phi^{++} &
-i\frac{\phi^+}{\sqrt{2}}\\-i\frac{\phi^+}{\sqrt{2}} &
\frac{-i\phi^0+\phi^P}{\sqrt{2}}
\end{array}
\right). \eeqn
Here, $H$ plays the role of the SM Higgs doublet, $h$
is the physical Higgs field and $v\simeq246$ GeV. The fields $\eta$
and $\omega$ are eaten by new heavy gauge bosons $A_H$, $W_H$ and
$Z_H$ when the $[SU(2) \times U(1)]^2$ gauge group is broken down to
$SU(2)_L\times U(1)_Y$, whereas the $\pi$ fields are absorbed by the
standard model $W/Z$ bosons after electroweak symmetry breaking
(EWSB). The field $h$ and $\Phi$ remain in the spectrum.
\par
Under the T-parity, the SM particles are T-even, and all the new
particles except $T_+$ (as we will introduce later) are T-odd. The
masses of T-even gauge bosons are given to $v^2/f^2$ order by:
\beqn
M_{W_L}=\frac{gv}{2}(1-\frac{v^2}{12f^2}),~~M_{Z_L}=\frac{M_{W_L}}{\cos
\theta_W},~~ M_{A_L} =0,
\eeqn
where $\theta_W$ is the weak mixing
angle, and $g$ is the SM $SU(2)$ gauge couplings.
\par
The masses of T-odd gauge bosons are given to $v^2/f^2$ order by:
\beqn M_{W_H}=gf(1-\frac{v^2}{8f^2}),~~M_{Z_H}=M_{W_H}, ~~M_{A_H}
=\frac{g'f}{\sqrt{5}}(1-\frac{5v^2}{8f^2})
\eeqn
where $g'$ is the
SM U(1) gauge couplings.

\subsection{Fermion sector}
The T-even fermion sector includes the SM quarks, leptons, and an
additional heavy quark $T_+$, which is introduced in the LH model in
order to cancel the quadratic divergence of the Higgs mass coming
from top loops. Each T-even fermion need a T-odd mirror fermion
under T-parity. The mirror quarks and leptons are involved in this
paper. We denote them by \beqn \left(\begin{array}{c} u_H^1 \\ d_H^1
\end{array} \right), ~~\left(\begin{array}{c} u_H^2
\\ d_H^2 \end{array} \right),~~\left(\begin{array}{c} u_H^3 \\
d_H^3 \end{array} \right).
\eeqn
and
\beqn
\left(\begin{array}{c} \nu_H^1 \\ l_H^1 \end{array} \right),
~~\left(\begin{array}{c} \nu_H^2
\\ l_H^2 \end{array} \right),~~\left(\begin{array}{c} \nu_H^3 \\
l_H^3 \end{array} \right).
\eeqn
\par
In ${\cal O}(v^2/f^2)$ their masses are given by
\beqn
&&m_{u_H^i}=\sqrt{2}\kappa_q^i f(1-\frac{v^2}{8f^2})=m_{Hi}(1-\frac{v^2}{8f^2}),
~~~ m_{d_H^i}=\sqrt{2}\kappa_q^if=m_{Hi} \\
&&m_{\nu_H^i}=\sqrt{2}\kappa_l^i
f(1-\frac{v^2}{8f^2})=m^l_{Hi}(1-\frac{v^2}{8f^2}), ~~~
m_{l_H^i}=\sqrt{2}\kappa_l^if= m^l_{Hi} \eeqn where $i$ is the
generation index, and $\kappa_q^i$ ($\kappa_l^i$) are the
eigenvalues of the mirror quark (lepton) Yukawa coupling matrices.
We neglect ${\cal O} (v^2/f^2)$ differences between $m_{u_H^i}$ and
$m_{d_H^i}$ ($m_{\nu_H^i}$ and $m_{l_H^i}$) in the numerical
calculation because these differences only contribute to higher
order corrections in the $v^2/f^2$ expansion.

\subsection{T-odd flavor mixing}
In the LHT model, the mirror fermions open up a new flavor structure
in the model. As discussed in
Ref.\cite{Hubisz:2005bd,Blanke:2006sb,Blanke:2006eb}, there are four
CKM-like unitary mixing matrices in the mirror fermion sector:
$V_{Hu}$, $V_{Hd}$, $V_{Hl}$ and $V_{H\nu}$. These mirror mixing
matrices are involved in the flavor changing interactions between SM
fermions and T-odd mirror fermions which are mediated by the T-odd
heavy gauge and Goldstone bosons ( $W_H,Z_H,A_H$ and
$\omega^\pm,\omega^0,\eta$), and they satisfy \beqn V_{Hu}^\dagger
V_{H{d}} = V_{\rm CKM},~~~~ V_{H\nu}^\dagger V_{Hl} = V^\dagger_{\rm
PMNS}, \eeqn
\par
Using the method in Ref.\cite{Blanke:2006eb,Blanke:2006xr}, $V_{Hd}$
is parameterized with three angles
$\theta_{12}^d,\theta_{23}^d,\theta_{13}^d$ and three phases
$\delta_{12}^d,\delta_{23}^d,\delta_{13}^d$, and analogously
$V_{Hl}$ is parameterized with three angles
$\theta_{12}^l,\theta_{23}^l,\theta_{13}^l$ and three phases
$\delta_{12}^l,\delta_{23}^l,\delta_{13}^l$. The explicit expression
won't be listed here. The Feynman rules for the flavor violating
interactions which are involved in our analysis can be found in
Ref.\cite{Blanke:2006eb,Blanke:2009am}.

\section{Calculation}
\par
\begin{figure}[htbp]
\centering
\scalebox{0.9}[0.9]{\includegraphics*[130pt,557pt][557pt,650pt]{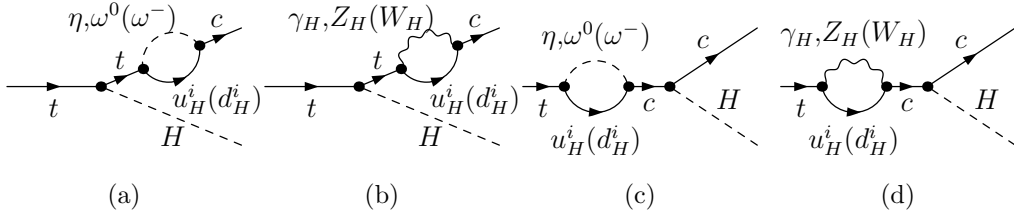}}
\caption{\label{feyn.tch} Feynman diagrams for \tch~ in the LHT
model.}
\end{figure}
\begin{figure}[htbp]
\centering
\scalebox{0.8}[0.8]{\includegraphics*[125pt,342pt][575pt,655pt]{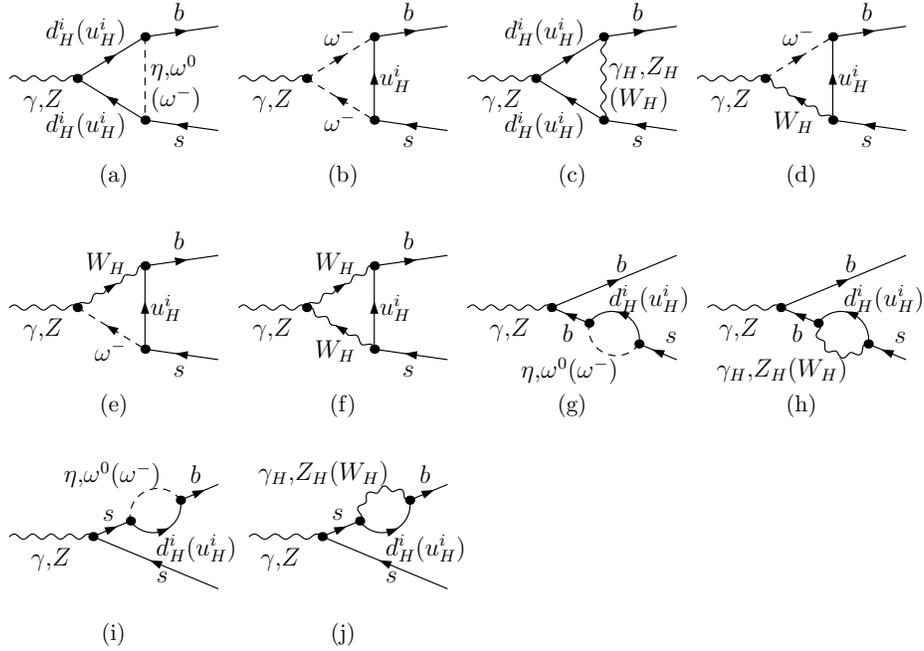}}
\caption{\label{feyn.zbs} Flavor changing vertex $\gamma(Z)b
\bar{s}$ in the LHT model.}
\end{figure}
\begin{figure}[!h]
\centering
\scalebox{0.8}[0.8]{\includegraphics*[125pt,232pt][575pt,650pt]{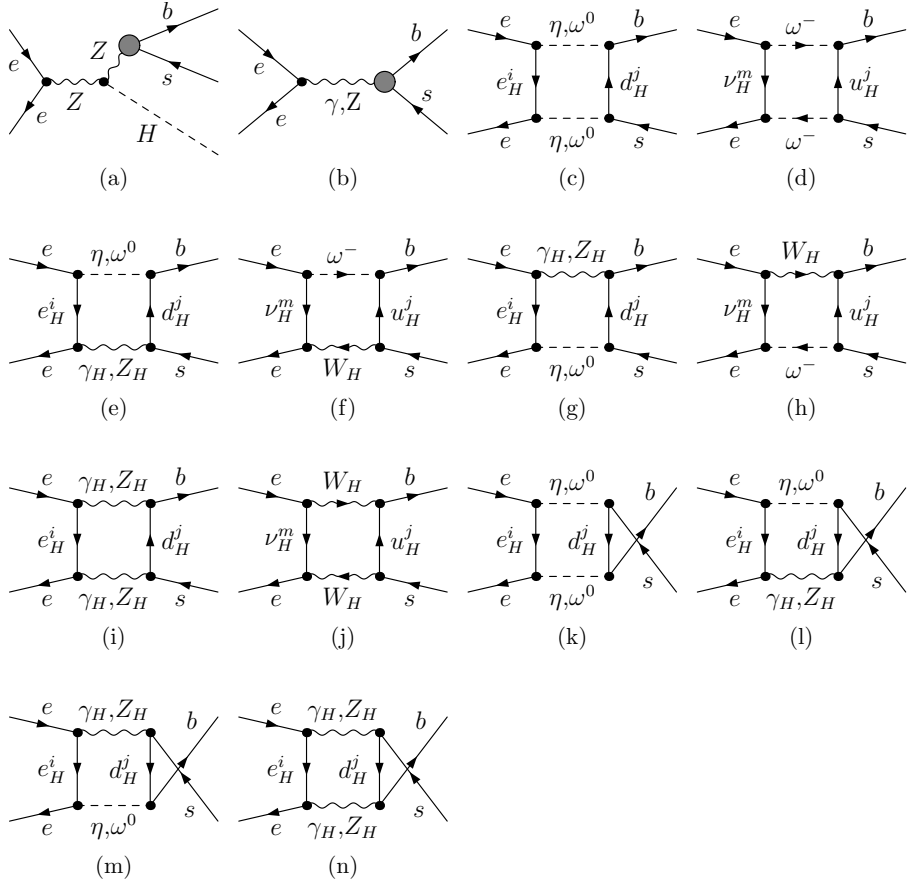}}
\caption{\label{feyn.eebs} One-loop Feynman diagrams for process
\eebsh ~and \eebs ~in the LHT model. The loop-induced
$\gamma(Z)b\bar{s}$ vertex (gray circle in (a) and (b)) is shown in
Fig.\ref{feyn.zbs}.}
\end{figure}

In this section we calculate the flavor changing effects originated
from heavy gauge bosons and mirror quarks on the processes \tch,
\eebs, \eebsh ~and \ppbs~ in the LHT model. The relevant Feynman
diagrams are shown in Figs.\ref{feyn.tch}-\ref{feyn.eebs}. The
diagrams for the subprocesses of \ppbs ~are similar to the diagrams
of process \eebs ~but much more, so we don't list them in the paper.
We have added the relevant Feynman rules of the LHT model to {\it
FeynArts3} package\cite{feynarts} and use it to generate the Feynman
diagrams and the corresponding amplitudes.  In the calculations of
the one-loop diagrams we adopt the definitions of one-loop integral
functions as in Ref.\cite{Passarino:1978jh}. The loop integral
functions are calculated by using the formulas in Ref.\cite{abcd}.

\par
Since there are no tree level diagrams for these processes except
for $pp \to u\bar{u}\to b\bar{s}$, we just sum all the
unrenormalized reducible and irreducible one-loop diagrams, and the
results will be finite and gauge-invariant. We checked these
processes ($pp \to u\bar{u}\to b\bar{s}$ not included) and found
that the divergences are canceled at \Ovdftwo~ for all the processes
except \tch. This divergence was explained as the sensitivity of the
decay amplitudes to the UV completion of the LH model in
Ref.\cite{Blanke:2006eb,Buras:2006wk}, and it is gauge independent.
We use the 't Hooft gauge in our calculation. After update some
vertices to \Ovdftwo~ in Ref.\cite{Blanke:2009am}, Goto et al. and
Blanke et al. found that the logarithmic divergence in Z boson
flavor changing processes can be canceled. We calculated the process
\tch ~with the updated Feynman rules, and found that the divergence
can't be canceled. So in our numerical calculations, we remove the
divergent term $1/\epsilon$ and take the renormalization scale
$\mu=\Lambda$ with $\Lambda=4\pi f$ being the cutoff scale of the
LHT model, as in Ref.\cite{Blanke:2006eb,Buras:2006wk}.

\par
In the numerical calculation, we take the SM parameters as follows
\cite{Nakamura:2010zzi,SMHiggs125GeV_ATLAS,SMHiggs125GeV_CMS}
\begin{eqnarray} \nonumber
&&\alpha=1/128,~ \alpha_s(m_Z)=0.1184,~ m_W=80.385~{\rm GeV},~
m_Z=91.1876~{\rm GeV},\\ \nonumber && \Gamma_Z=2.4952~{\rm GeV},
~m_t=173.5~{\rm GeV}, ~m_c=1.27~{\rm GeV},
~m_b=4.67~{\rm GeV}, \\
&& m_s=0.101~{\rm GeV}, ~m_H=125~{\rm GeV}.
\end{eqnarray}

For the scattering processes, we take the cuts on final particles as
$p_T^{b,s,H} \ge 15~{\rm GeV}$.

Considering the constraints on PMNS matrix\cite{Mena:2003ug,
Mohapatra:2005wg, Ahuja:2006yf, DayaBay}, we set PMNS parameters to
\beqn
s_{12}=\sqrt{0.3},~s_{23}=\sqrt{0.5},~s_{13}=\sqrt{0.024},~\delta=65^\circ,
\eeqn
where the Majorana phases in $V_{PMNS}$ have been set to zero,
because no Majorana mass term has been introduced for right-handed
neutrinos.

\par
The LHT parameters which are relevant to our analysis are
\beqn
f,~ m_{H1},~m_{H2},~m_{H3},~\theta_{12}^d,~\theta_{23}^d,~\theta_{13}^d,
~\delta_{12}^d,~\delta_{23}^d,~\delta_{13}^d \\
\label{lepton}
 m^l_{H1},~m^l_{H2},~m^l_{H3},~\theta_{12}^l,~\theta_{23}^l,~\theta_{13}^l,
~\delta_{12}^l,~\delta_{23}^l,~\delta_{13}^l
\eeqn

The LHT scale $f$ can be as low as 500 GeV\cite{Hubisz:2005tx}, so
we vary it in the range $500~ {\rm GeV}\leq f \leq 1500~ {\rm GeV}$.
The constrains on the mass spectrum of the mirror fermions have been
extensively studied
\cite{Hubisz:2005bd,Blanke:2006sb,Blanke:2006eb,Blanke:2009am,Blanke:2007ee,Blanke:2008ac,Bigi:2009df,Blum:2009sk}.
It is convenient to consider several representative scenarios for
the structure of the matrix $V_{Hd}$. In
Ref\cite{Blanke:2006sb,Blanke:2006eb} several benchmark scenarios
was introduced, among which Scenario 4 allows for large effects in
the $B_s$ system. In this scenario the hierarchical structure of
$V_{Hd}$ matrix is very different from the structure of the CKM
matrix, and they assume that
\begin{eqnarray}
\label{scenario4} && ~ \frac{1}{\sqrt{2}}\leq s^d_{12}\leq0.99,
 ~ 5\times 10^{-5}\leq s^d_{23}\leq 2\times10^{-4},
 ~ 4\times 10^{-2}\leq s^d_{13}\leq 0.6.
\end{eqnarray}
To be simplicity we choose the lower and upper limits of $s^d_{ij}$
($(ij)=(12),(23),(13)$) in Eq.(\ref{scenario4}) as Case II and Case
III, respectively, with the phase term $\delta^d_{ij}=0$. In Case I
we assume that there are no mixing in down type mirror quarks, i.e.,
$V_{Hd}={\rm 1}$. We follow Ref\cite{Han:2008wb} to give the values
of mirror fermions. Here we list the there cases we used,

\begin{itemize}
\item {\bf Case I,}
$V_{Hd} = {\mathbf 1}$
\item {\bf Case II,}
 $s^d_{12}=\frac{1}{\sqrt{2}},
 ~s^d_{23}=5\times 10^{-5},
 ~s^d_{13}=4\times 10^{-2},
 ~\delta^d_{12}=\delta^d_{23}=\delta^d_{13}=0,$
\item {\bf Case III,}
 $s^d_{12}=0.99,
 ~ s^d_{23} = 2\times10^{-4},
 ~ s^d_{13} = 0.6,
 ~ \delta^d_{12}=\delta^d_{23}=\delta^d_{13}=0$
\end{itemize}

\par
In all the three cases the first two mirror quark generations are
chosen to be quasi-generate. In Case II and III we take the masses
of mirror fermions as
\begin{eqnarray}
&& m_{d^1_H}=m_{d^2_H}=\frac{600 {\rm ~GeV}}{\rm TeV} f, ~
m_{d^3_H}=\frac{1400 {\rm ~GeV}}{\rm TeV} f.
\end{eqnarray}

\par
The lepton sector in Eq.(\ref{lepton}) is only involved in the
process \eebs. According to Scenario C in Ref.\cite{Blanke:2007db},
we constrain the mirror lepton masses to lie in the range $300~{\rm
GeV} \le m_{Hi}^l \le 1.5~{\rm TeV}$, and scan over the whole
parameter space for the mirror lepton mixing part, i.e. $0 \le
\theta_{ij}^l \le 2\pi$ and $0 \le \delta_{ij}^l \le 2\pi$.

\par
For the decay process \tch, the branching ratio is defined as follow
because $t \to b W^+$ is the dominant channel of top-quark
\beqn
Br(t \rightarrow cH)=\frac{\Gamma(t \rightarrow c
H)}{\Gamma(t\rightarrow bW^+)}.
\eeqn
For the other processes we
show the cross sections as final results.

\par
Case I only contribute to \tch ~and $pp \to u\bar{u} \to b\bar{s}$
processes, however the cross section for the latter is so small
($\sim 10^{-7}~{\rm fb}$) that can be neglected. The other processes
have no contribution from mirror quark loops, because there are no
flavor mixing between down-type mirror quarks in Case I.
\begin{figure} \centering
\scalebox{0.4}[0.4]{\includegraphics*[52pt,16pt][580pt,413pt]{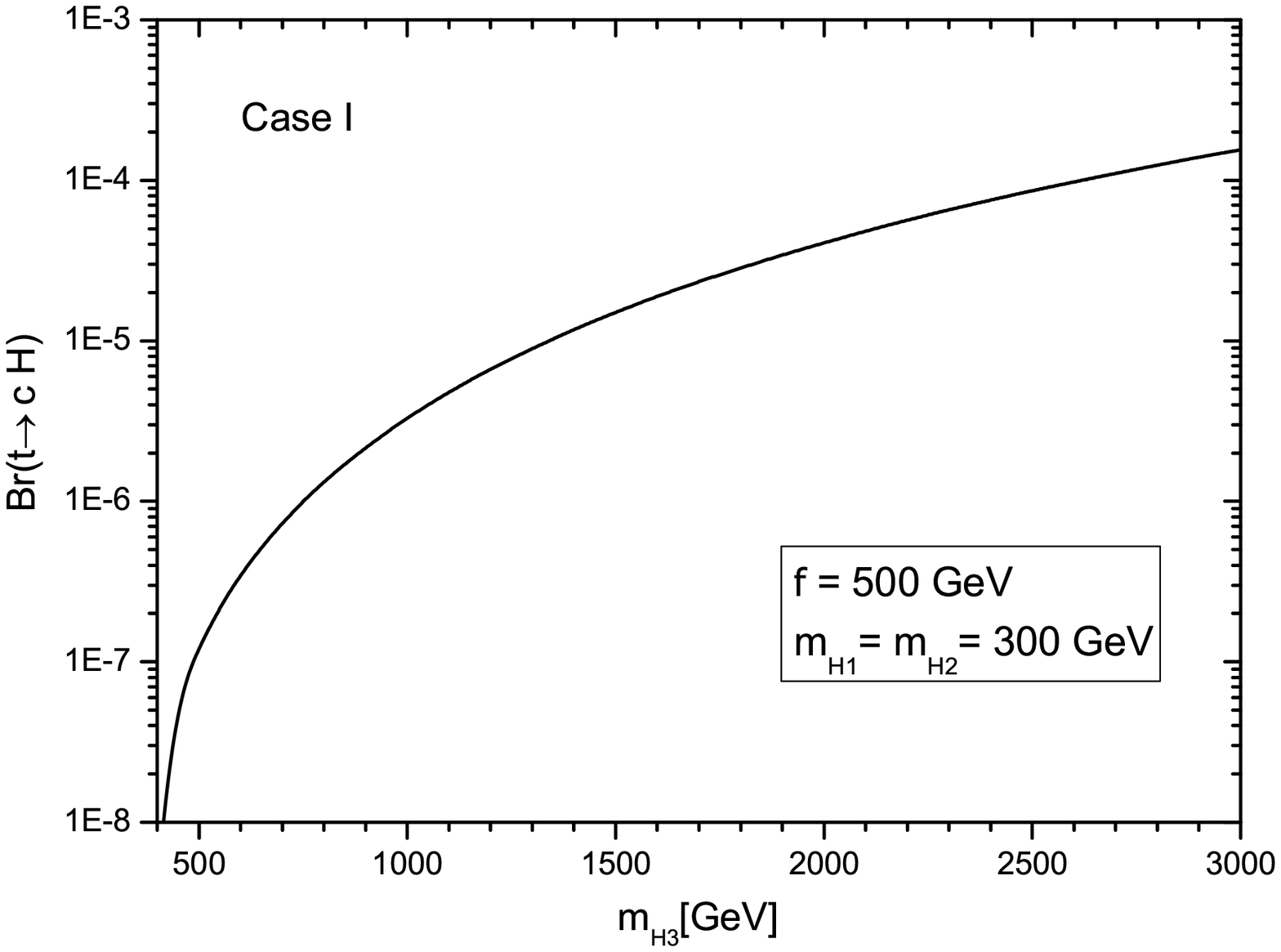}}
\scalebox{0.4}[0.4]{\includegraphics*[48pt,16pt][580pt,413pt]{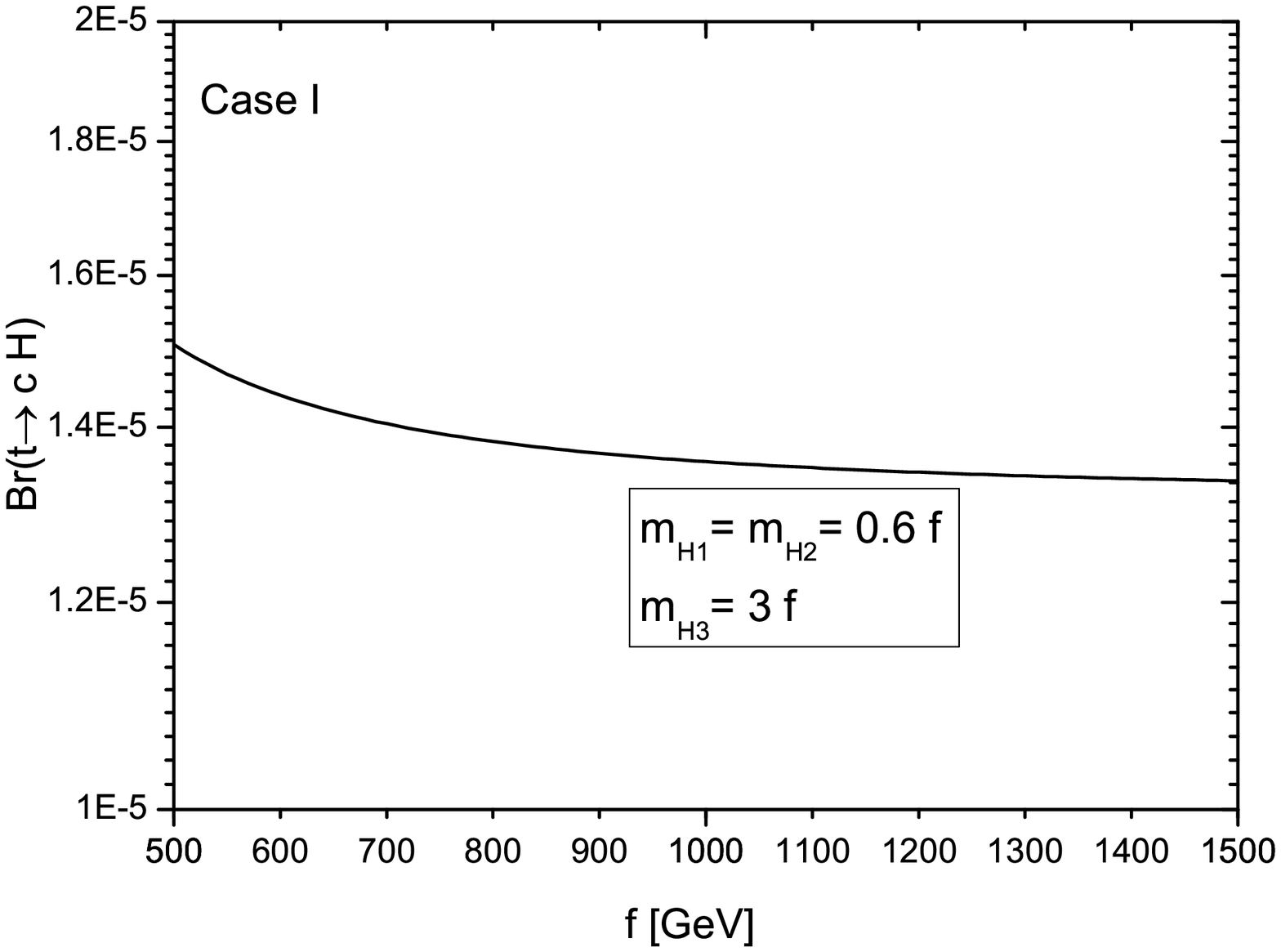}}
\caption{\label{tch} Branching ratios for \tch ~ in Case I, as the
function of $m_{H3}$ (left) and $f$ (right).}
\end{figure}

\par
In Fig.\ref{tch} we show the branching ratios of \tch ~decay process
as the functions of the mass of the third generation mirror quark
(left) and the LHT scale $f$ (right) in Case I. We set
$m_{H1}=m_{H2}=300~{\rm GeV}$ and $f = 500~{\rm GeV}$ in the left
figure. The branching ratio increases with $m_{H3}$, because the
decay rate is enhanced by the mass splitting between the three
generation mirror quarks, and since we set $m_{H1}=m_{H2}$, there is
only one mass splitting $m_{H3}-m_{H2}$, which increases with
$m_{H3}$. In the right figure we set $m_{H1}=m_{H2}=0.6 f$ and
$m_{H3}=3 f$, and we can see that the branching ratio decreases with
$f$, but very slowly. That's because the mass splitting
$m_{H3}-m_{H2}$ is large enough ($2.4 f$) to cancel the the
decrement caused by large $f$. We can also see that the branching
ratio enhances a lot in the LHT model compared with that in the SM
($10^{-14}$). At the same time the contributions from the LHT model
are much larger than those in the LH model without T-parity ($\sim
10^{-12}$\cite{Tabbakh:2005kf}). That's beacuse the parameters have
less constraints, and the mirror fermions in the loops contribute a
lot. The branching ratio can reach $10^{-4}$ when $f=500~{\rm GeV}$,
$m_{H1}=m_{H2}=300~{\rm GeV}$ and $m_{H3}=3~{\rm TeV}$, which is
even larger than the SUSY-QCD contribution\cite{tch.mssm}.
\begin{figure}
\centering
\scalebox{0.4}[0.4]{\includegraphics*[48pt,16pt][580pt,413pt]{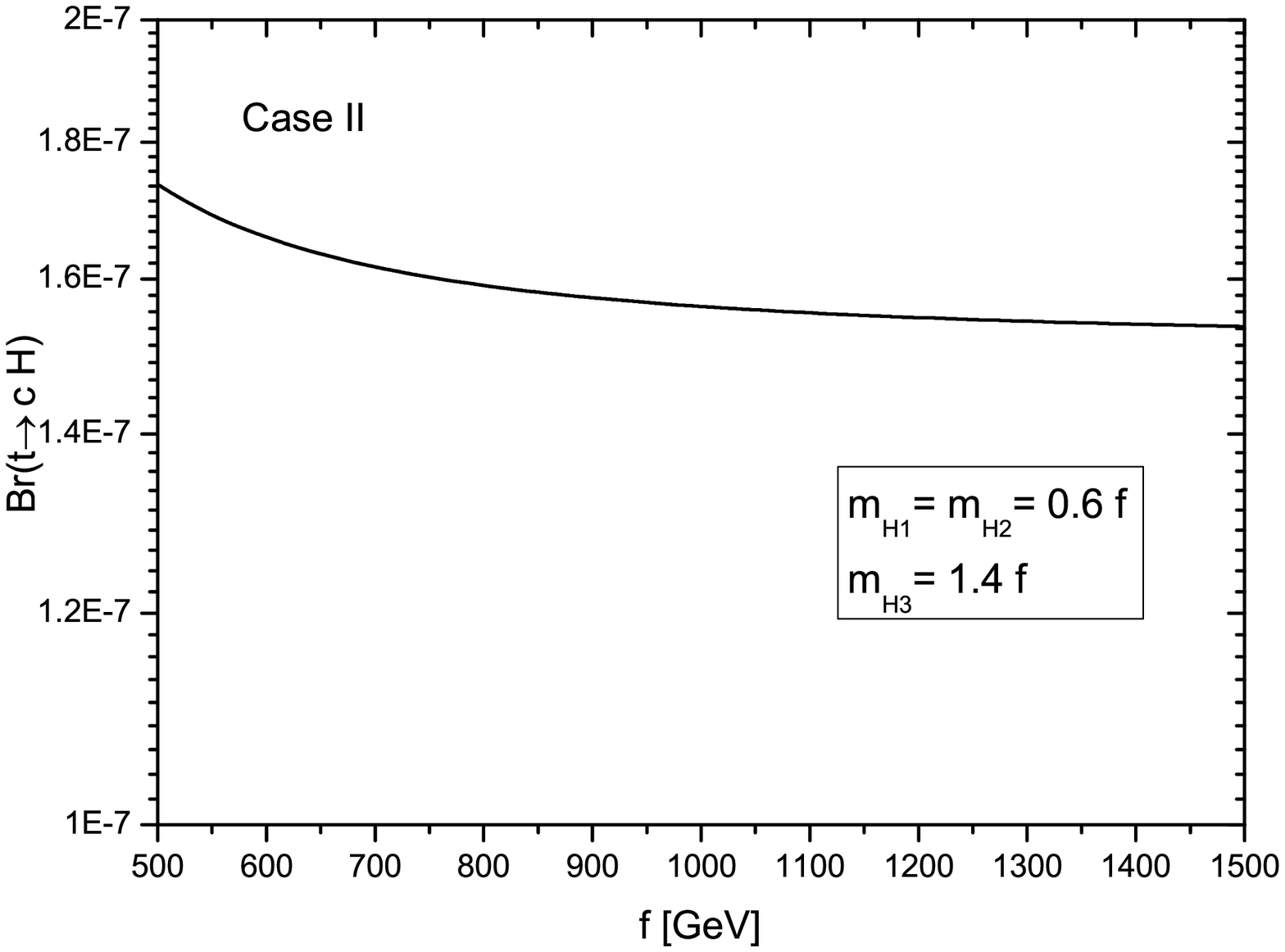}}
\scalebox{0.4}[0.4]{\includegraphics*[48pt,16pt][580pt,413pt]{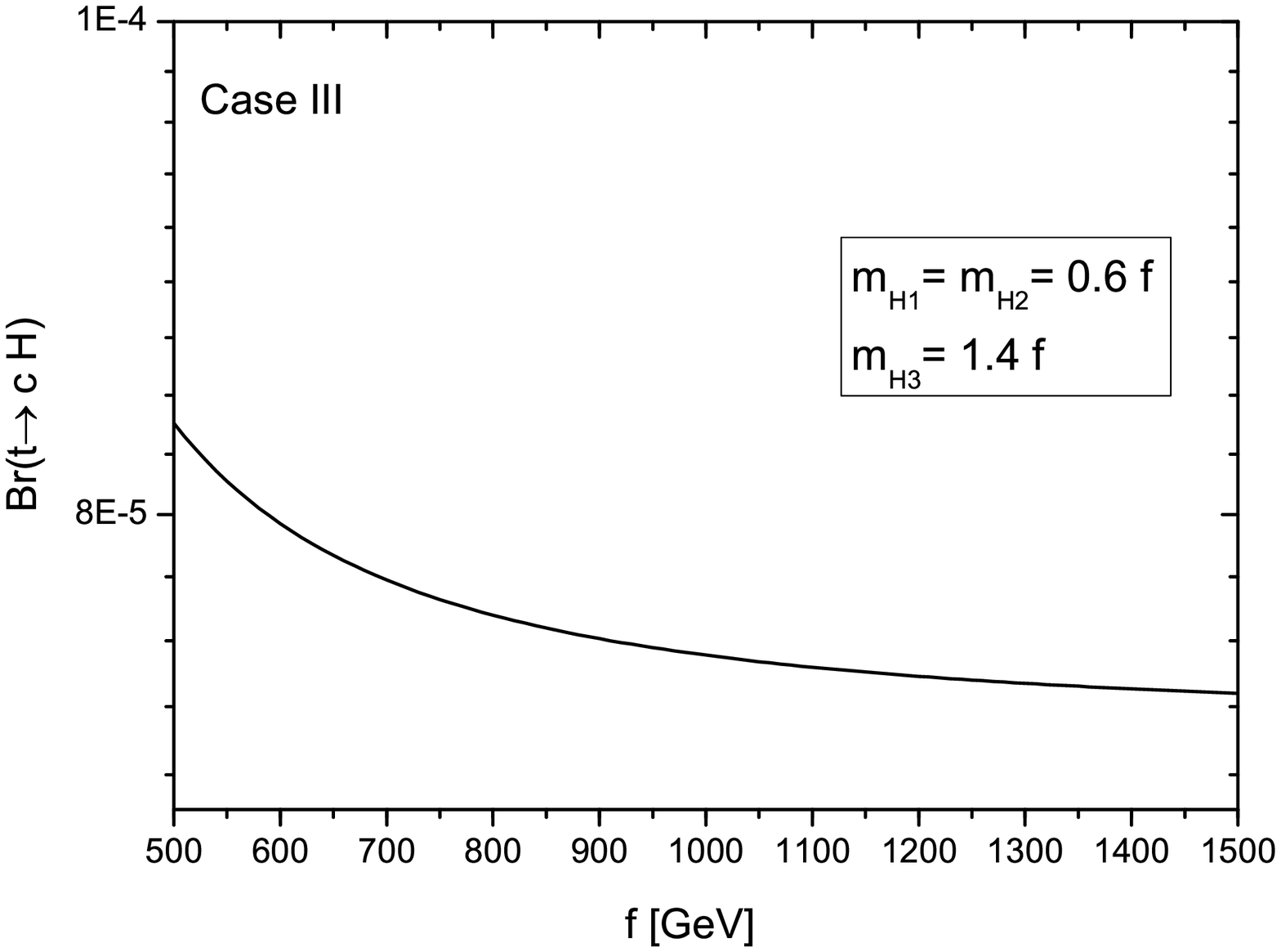}}
\caption{\label{tch2} Branching ratios for \tch ~as the function of
$f$ in Case II and III.}
\end{figure}

In Fig.\ref{tch2} we present the branching ratio of \tch ~process as
the function of $f$ in Case II and III. In both cases the branching
ratios decease with $f$ but very slowly, for the same reason stated
above. The branching ratio in Case III is much larger than that in
Case II, because the mixing between mirror quarks is much larger.

\begin{figure}[h]
\centering
\scalebox{0.4}[0.4]{\includegraphics*[52pt,0pt][580pt,413pt]{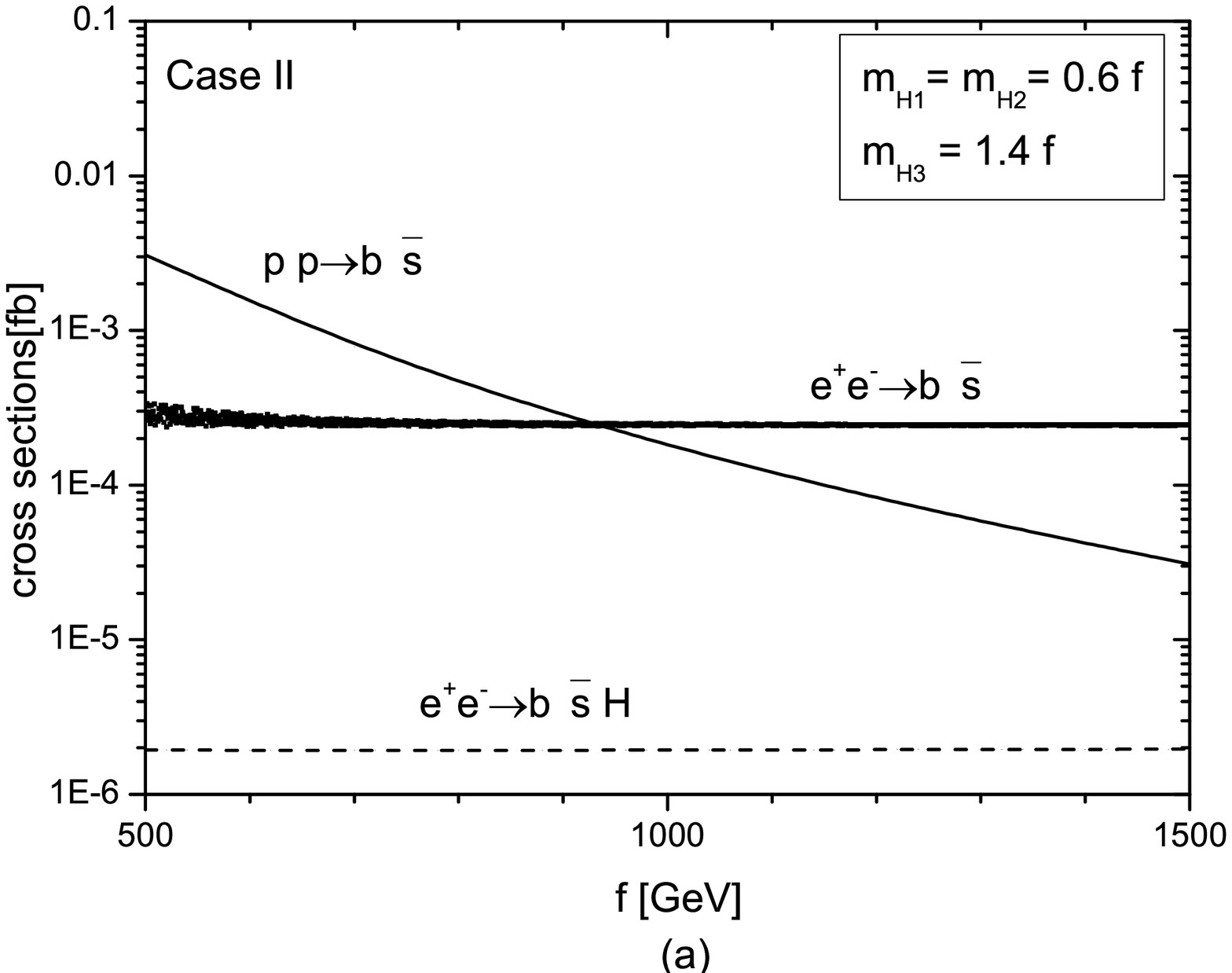}}
\scalebox{0.4}[0.4]{\includegraphics*[52pt,0pt][580pt,413pt]{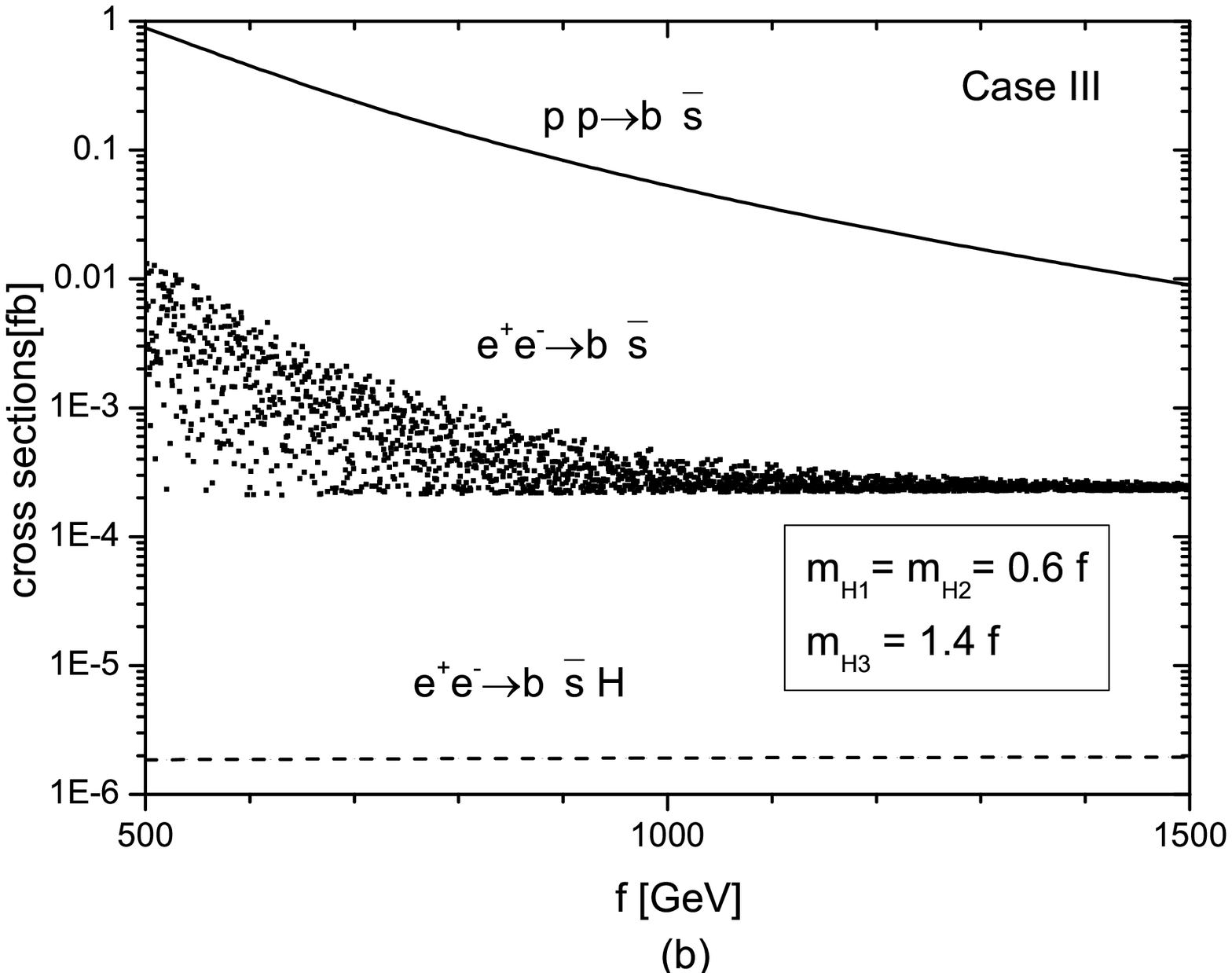}}
\caption{\label{caseII} Cross sections for \ppbs ~at the 14 TeV LHC,
\eebs ~and \eebsh ~at a 500 GeV linear collider, as the functions of
$f$ in Case II (left) and III (right).}
\end{figure}
\par
In Fig.\ref{caseII} we present the cross sections for processes
\eebs, \eebsh ~and \ppbs ~as the functions of $f$ for Case II (left)
and III (right). The center of mass system (c.m.s.) energy
$\sqrt{s}$ is 14 TeV at LHC and 500 GeV at a linear collider. We
scanned the mirror lepton sector parameters over the whole space
except the restriction for mirror lepton masses when computing the
\eebs ~cross sections. In Case II the cross sections for \eebs~
process in the LHT model almost degenerate with those in the SM in
most of the parameter space, only a small enhancement in the low $f$
region. While in Case III the LHT effect could enhance the SM cross
section by two orders, with $f$ close to 500 GeV. In both cases the
LHT effects on \eebsh~ process are very small, only enhance the SM
cross section by a few percent.

Coming to \ppbs ~process, there is a tree-level diagram of flavor
changing charged current for the subprocess $u\bar{u}\to b\bar{s}$.
If we compute one-loop contribution, there would exist UV and IR
divergences, so the renormalization procedure is necessary. For
simplicity we estimate the cross section in the SM, and compute the
pure LHT effect on this process at 1-loop level. This means that we
set the CKM matrix to be unit when we generate the Feynman diagrams
and amplitudes in the LHT model, and in this case there are no tree
level diagrams, so we can sum all the diagrams together and the
result would be finite. Now let's make a rough comparison between
the cross section in the SM and in the LHT model. First we list the
cross sections for the three subprocesses at LHC:
\begin{eqnarray}
\sigma^{tree}_{SM}(pp\to u\bar{u}\to b\bar{s})=0.449~ {\rm fb} \\
\sigma^{1-loop}_{SM}(pp\to d\bar{d}\to b\bar{s})=0.505~ {\rm fb} \\
\sigma^{1-loop}_{SM}(pp\to gg\to b\bar{s})=0.604~ {\rm fb}
\end{eqnarray}
Supposing the QCD correction of $pp\to u\bar{u}\to
b\bar{s}$ to be $20\%$, we got \beqn \sigma^{1-loop}_{SM}(pp\to
u\bar{u}\to b\bar{s})\sim 0.449*0.2~ fb =0.898~fb
\end{eqnarray}
Summing the tree level and 1-loop level subprocess cross sections
together, we obtain the total cross section of \ppbs ~process in the
SM at 1-loop level
\begin{eqnarray} \sigma^{1-loop}_{SM}(pp\to
b\bar{s})\sim 1.65~fb
\end{eqnarray}
From Fig.\ref{caseII}(a) we can see that the pure LHT cross section
of \ppbs ~varies from $3\times 10^{-5} $ to $3\times 10^{-3} $ fb,
which is much smaller than the SM contribution. In
Fig.\ref{caseII}(b) the pure LHT cross section for \ppbs ~enhances 2
orders compare with Fig.\ref{caseII}(a), but still smaller than the
SM cross section, and can't be detected at LHC.

\begin{figure}[!h]
\centering
\scalebox{0.5}[0.5]{\includegraphics*[48pt,20pt][578pt,418pt]{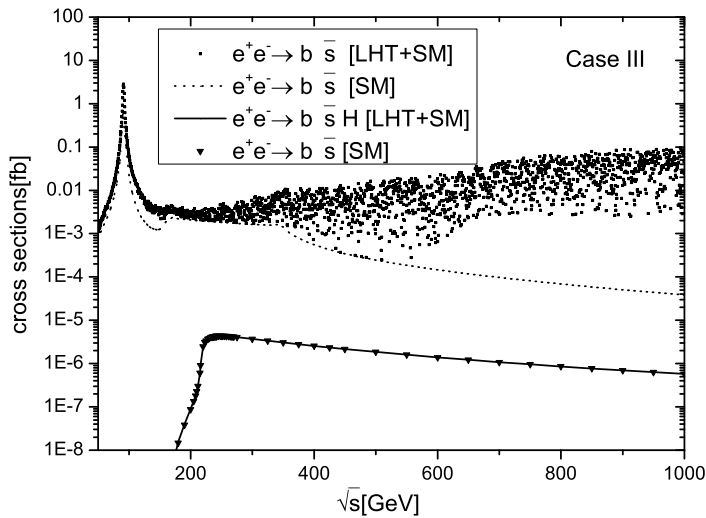}}
\caption{\label{runS} Cross sections for \eebs ~and \eebsh ~as the
functions of $\sqrt{s}$ in Case III, with  $f=500~{\rm GeV}$,
$m_{H1}=m_{H2}=300~{\rm GeV}$ and $m_{H3}=700~{\rm GeV}$.}
\end{figure}

In Fig.\ref{runS} we present the cross sections for \eebs ~and
\eebsh ~as the functions of $\sqrt{s}$ in Case III, with $f=500~{\rm
GeV}$, $m_{H1}=m_{H2}=300~{\rm GeV}$ and $m_{H3}=700~{\rm GeV}$.
There are three peaks in the curves for \eebs~ process,
corresponding to the resonance of Z boson, a pair of W boson
threshold, and a pair of top quark threshold, respectively. The
cross section for \eebsh ~decreases with the increase of $\sqrt{s}$
beyond the Higgs resonance peak (250 GeV) because of the s-channel
depression. We can also see that the LHT effect on \eebs~ process
increase with $\sqrt{s}$, and could enhance the SM cross section by
3 orders when $\sqrt{s} \gsim 800 {\rm GeV}$. The cross section can
reach 0.1 fb with large $\sqrt{s}$, and even exceed 1 fb at Z
resonance. So it might be possible to see the LHT effect on this
process at future linear colliders. While the cross section in the
LHT model for \eebsh~ process almost degenerates with that in the
SM, and they are too small to be detected at the future linear
colliders.

\section{Summary}
In this paper we calculate the one-loop contributions from heavy
gauge bosons and mirror fermions to the top quark rare decay process
\tch, $b\bar{s}(H)$ production at linear colliders, and $b\bar{s}$
production at the 14 TeV LHC in the LHT model. The branching ratio
for \tch~ can reach $10^{-4}$ in the parameter space we considered,
which is much larger than that in the SM, and could be detected in
the experiment. With a relative small $f$ value ($\sim$ 500 GeV) and
large $\sqrt{s}$ ($\sim$ 1 TeV), the LHT could enhance the SM cross
section by three orders and reach 0.1 fb, which might be possible to
be seen at future linear colliders. While the LHT have much smaller
effect on process \eebsh~ and \ppbs~ thus couldn't be detected at
future linear colliders and LHT.

\section{Acknowledgments}
Project supported by the National Natural Science Foundation of
China (Grant No.11105083, No.10947139, No.11035003, No.10947022,
No.11105036, No.11147151 and No.11205070), and by Shandong Province
Natural Science Foundation (No.ZR2012AQ017)

\end{document}